# Understanding the Maker Protocol[1]

Jason Chen
Vandegrift High School
216 Summer Alcove Way
Austin, TX 78732

Kathy Fogel
School of Management
Clark University
Worcester, MA 01610

Kose John
Stern School of Business
New York University
New York, NY 10012

**October 29, 2022**

Abstract:

This paper discusses a decentralized finance (DeFi) application called MakerDAO. The Maker Protocol, built on the Ethereum blockchain, enables users to create and hold currency. Current elements of the Maker Protocol are the Dai stable coin, Maker Vaults, and Voting. MakerDAO governs the Maker Protocol by deciding on key parameters (e.g., stability fees, collateral types and rates, etc.) through the voting power of Maker (MKR) holders. The Maker Protocol is one of the largest decentralized applications (DApps) on the Ethereum blockchain and is the first decentralized finance (DeFi) application to earn significant adoption. The objective of this paper is to analyze and discuss the significance, uses, and functions of this DeFi application.

---

1. **INTRODUCTION**

In April 2020, as the pandemic shook global markets, bitcoin traded at a price below $7,000. One and a half years later, in November 2021, bitcoin's price increased nearly tenfold to $67,000. Then, on September 1, 2022, the market price of bitcoin fell again, this time to $19,734. Many cryptocurrencies experience similar price volatility. This high volatility limits their functionality as a medium of exchange.[2] Stable coins address this weakness. Stable coins are cryptocurrencies that seek low price volatility by aiming to trade in a narrow band relative to some other "stable" asset such as the U.S. dollar.

Decentralized finance, or DeFi, is a variety of financial applications that use blockchains to replace traditional intermediaries. These intermediaries include banks, insurance companies, brokerages, and the like. While some big blockchain users, such as Bitcoin, are used more as a transactions cryptocurrency, many other DeFi applications use the blockchain platform for non-custodial services such as lending. In this paper, we will be discussing the DeFi application (DApp) known as Maker. The cryptocurrency Dai is among the most prominent stable coins. This stable coin operates under the Maker Protocol, which, in turn, is governed by the decentralized autonomous organization MakerDAO.

2. **DIGGING INTO MAKER**

[Insert Figure 1 here.]

---

[2] In addition to price volatility the payments function of many cryptocurrencies, including bitcoin, is further restricted by the economic implications of their protocol design. For more detailed reference see Hinzen, John and Saleh (2022).

A blockchain is an electronic ledger that records data in discrete chunks called "blocks." Most well-known blockchains are public networks. This means that the contents of the ledger can be read by anyone.[3] Further, they are permissionless, which means that they are jointly maintained by a set of distinct participants and that no permissions are required for becoming such a participant. While early blockchains such as the Bitcoin were limited in the type of data stored on the ledger, modern blockchains allow for the storage of code and its automatic execution. Such pieces of code are called *smart contracts*.[4] The most commonly used blockchain that allows for smart contracts is Ethereum. Some smart contracts set out rules, called protocols, under which further cryptocurrencies are created. Maker is a protocol built on the Ethereum blockchain.

The purpose of the Maker Protocol is to generate the stable coin Dai, which aims to function as a comprehensive medium of exchange. The price of Dai is targeted to trade within a narrow band with respect to the U.S. dollar. Maker aims to achieve price stability of Dai relative to this base currency via its governance token MKR. MKR is an ERC-20 token (a standard used for creating and issuing smart contracts on the Ethereum blockchain) native to the Maker Protocol. Unlike Dai, MKR's price is not bound to another asset and may therefore be more volatile. MKR acts as a utility token and is needed for paying the fees accrued on Collateralized Debt Positions, or CDPs for short. The CDP, MakerDAO's smart contract loans are used to generate Dai in the Maker system. Only MKR can pay these fees, and when paid, the MKR is burned, removing it from the supply. This means that if the adoption and demand for Dai and CDPs increase, there will be additional demand for MKR so that users can pay the fees. In the same vein, the supply will decrease as MKR is burned. In addition, MKR holders gain voting rights proportional to the amount of MKR they hold. MKR holders vote on features of MakerDAO, an example being the amount of collateral needed for a CDP. Participating in these votes isn't simply a way for users to influence the platform; voting gives rewards in the form of additional MKR tokens. These tokens can then be resold for profit or held and used to gain greater influence over MKR votes.

---

[3] For a comprehensive introduction to the technology of crypto currencies, see Narayanan, Bonneau, Felten, Miller, and Goldfeder (2016). For a recent survey on Bitcoin and other cryptocurrencies, see John, O'Hara and Saleh (2022).
[4] For a recent survey on "Smart Contracts and DeFi," see John, Kogan and Saleh (2022).

[Insert Figure 2 here.]

Dai is created by opening CDP, which is the position resulting from locking collateral in MakerDAO's smart contract. For a user to obtain a Dai coin, she will need to deposit Ether (or any other cryptocurrency that can be accepted as collateral) in order to borrow against their deposits and receive the newly generated Dai. To understand how this works, it's necessary to understand how MakerDAO's special type of loan (CDPs) works. CDPs are bought with Ether, and, in exchange, a user receives Dai. The Ether that is paid acts as collateral for the loan. In other words, a user can take out a loan against her Ether holdings. Upon repaying the loan, the issued Dai is destroyed. There are also MKR transaction fees during this process. If the price of Dai is too low, then MKR can be produced and sold on the market to raise additional collateral and cover the debt obligation.

CDPs can be thought of as secure vaults for storing the aforementioned collateral. To account for the volatility in the crypto collateral, Dai is often over-collateralized, meaning that the deposit amount required is typically higher than the value of Dai. For example, users must spend $200 in Ether in order to receive $100 Dai, which is meant to account for the potential decrease in the value of Ether. As a result, if Ether depreciates by 25%, the $100 in Dai would still be safely collateralized by $150 in Ether. The minimum collateral ratio for Ether is currently 150%, meaning that if $150 worth of Ether is deposited by a user, then they would receive about 100 Dai (approximately $100).

However, the price of Ether varies freely. Thus, in principle it is possible that the value of the collateralizing Ether may fall below its stipulated minimum. To ensure there is enough collateral in the Maker Protocol to cover the value of all outstanding debt (the amount of Dai outstanding valued at its $1 peg), any Maker Vault deemed too risky (according to parameters established by Maker Governance) is liquidated through automated Maker Protocol auctions. The Protocol makes such decisions automatically after comparing the liquidation ratio (the collateral-to-debt ratio at which a Vault becomes vulnerable to liquidation) to the vault's current collateral-to-debt ratio. Each Vault typically has its own values of the liquidation ratio, and each ratio is always determined by the MKR voters based on the risk profiles of collateral asset types.

3. **TOTAL VALUE LOCKED**

The Total Value Locked, or TVL, is the combined overall value of crypto assets deposited in the DeFi protocol. It is used as a key metric for gauging interest in protocol sectors of the crypto industry. TVL includes all functions that DeFi protocols offer, including staking, lending, and liquidity pools. Importantly, it reflects the current value of the deposits themselves rather than the yield that these deposits are expected to earn. Investors can look at TVL when assessing whether a DeFi project's native token is valued appropriately. The market cap of the token may be high or low relative to the TVL of the project. The more extreme the relationship, the more overvalued or undervalued the token may appear.

[Insert Figure 3 here.]

Since the creation of the Maker Protocol, its TVL has changed drastically since its release. The first version of MakerDAO and Dai white paper was published in December 2017, and thus the first stable coin governed by a DAO (decentralized autonomous organization) in the entire crypto world was born. The doors were opened to a new range of financial opportunities on blockchain technology. Subsequently, in September 2018 the foundation proposal was approved and the following year Dai became live. Dai can now be generated from various crypto assets approved by MKR holders, and, in September 2020, the amount of generated Dai's reached half a billion. In the following two years, the protocol gained critical value until February 10, 2022. On that day, Maker's TVL was roughly $18.2 billion, almost 45% of its year-high; By May 24, 2022, it fell to $9.82 billion, and then, on September 21, 2022, sank to $7.26 billion.

[Insert Figure 4 here.]

MakerDAO's TVL fell substantially due to an overall crypto market crash. Maker TVL was $17.5 billion on the first day of this year, 2022. The TVL declined to less than $9 billion in October. In comparison, Ethereum TVL fell from $146.78 billion on January 1, 2022 to as low as $52.99 billion on September 16, 2022.

[Insert Figure 5 here.]

As of October 26th, Makers TVL is worth around $8.2 billion, and it is slowly climbing back. Despite shedding more than $9 billion in TVL, MakerDAO has become the protocol on Ethereum with the most value locked. In other words, MakerDAO holds a relatively higher TVL than other popular protocols such as Curve, Aave, Lido, Uniswap, Convex Finance, PancakeSwap, Compound, JustLend, Instadapp, and SushiSwap.

4. **MakerDAO APPLICATIONS**

Every Dai in circulation is directly backed by excess collateral, meaning that the value of the collateral is higher than the value of the Dai debt, and all Dai transactions are publicly viewable on the Ethereum blockchain. The smart contracts that MakerDAO uses are also called Maker Vaults. Users can access the Maker Protocol and create Vaults through a number of different user interfaces such as Oasis Borrow, MyEtherWallet, or Zerion. Vaults are inherently non-custodial: users interact with Vaults and the Maker Protocol directly, and each user has complete and independent control over their deposited collateral as long the value of that collateral doesn't fall below the required minimum level. Creating a vault is simple, but generating the Dai creates the obligation to repay the Dai along with a stability fee to withdraw all the collateral within a Maker Vault.

The stability fee is a risk parameter designed to address the inherent risk in generating Dai against collaterals stored in Maker Vaults. A part of the stability fee is also set aside for the purpose of sustaining operations of the Maker Protocol which includes the Dai Savings Rate (DSR), and other costs inherent to the protocol. The DSR is a variable rate of accrual earned by locking Dai in the DSR smart contract. Dai holders can earn savings automatically and natively while retaining control of their Dai. The DSR smart contract has no withdrawal limits, deposit limits, or liquidity constraints. The rate is actively set by MKR token holders through on-chain governance.

5. **INTERACTING WITH MAKER VAULTS**

A user can create a vault, preferably through Oaisis Borrow, by funding with a specific amount of collaterals that can be used to generate Dai. Once funded, the vault can then be considered collateralized. After creating a vault, the user must initiate a transaction and then confirm it in their unposted cryptocurrency wallet in order to generate a specific amount of Dai in exchange for keeping her collateral locked in the Vault. If a user wishes to withdraw a portion or all collateral assets, the user must pay back the exact amount in Dai that they generated, in addition to paying a stability fee on the outstanding Dai. The stability fee can only be paid in Dai. After a user returns their Dai and pays the stability fee, she can now withdraw some or all of her collateral. If all the Dai is returned, and all the collateral is retrieved, the vault will close and stay empty until the user decides to make another deposit. Each collateral asset deposited requires its own vault. Some users will own multiple vaults with different types of collateral and levels of collateralization.

### 6. MAKER VERSUS CENTRALIZED FINANCE

The Federal Reserve System is the centralized banking system of the United States. In a centralized system, a bank's primary function is to lend account holders' money to other people, who will in turn use that money to buy homes, businesses, pay back debt, and the like. When someone deposits money into a bank account, that money goes into a large pool, and the amount of money that is deposited is credited to that person's account. Money is subtracted from the account when a check is written, or withdrawals are made. The Federal Reserve Bank prints money and issues coins to release into the circulation; they also set the reserve requirements for banks. Its Board of Governors sets interest rates and approves changes to the banking system, and its Open Market Committee sets monetary policy.

The Maker protocol aims to function as a comprehensive medium of exchange, in almost the same way that the Federal Reserve systems do. Maker is made up of a smart contract service that manages borrowing and lending, as well as two currencies: Dai and MKR to regulate the value of loans. The Maker Protocol is managed by users around the world who hold its governance token MKR, and Dai stable coin is its decentralized, unbiased, collateral-backed cryptocurrency soft-pegged to the US Dollar. Users can generate Dai by depositing collateral assets into Maker Vaults within the Maker Protocol.

[Insert Figure 6 here.]

Despite the two systems being very similar in terms of purpose, there are important differences that put Maker on top. Maker is a decentralized protocol built on the Ethereum network to allow lending and borrowing without the need for an intermediary, a third party that facilitates transactions between two possibly unrelated parties. A bank is an intermediary for loans because consumers deposit money with the bank and banks loan out that money to other consumers for mortgages. The Maker Protocol does not require intermediaries, and users have full control over all their assets, meaning Maker cannot loan assets of users to other users. In the case of a sudden economic crisis, the maker protocol will initiate an emergency shutdown. Examples of economic crises include a severely broken peg (the linking of the market value of a cryptocurrency to an external reference, which can be a fiat currency or a commodity) and other market scenarios if they pose a significant and real threat to the majority of users. Maker's emergency shutdown locks the system, stopping Vault creation and the ability to generate Dai and locally freezing the Reference Prices. Vault owners can immediately withdraw excess collateral. The Federal Reserve provides key financial services to the nation's payment system including distributing the nation's cash and coin to banks, as well as adjusting interest rates. The Federal Reserve however does struggle more when dealing with economic crises. For example, during the Great Depression, the Federal Reserve raised interest rates too much, and borrowers were forced into default, resulting in panic and a stock market crash. Everyone wanted to withdraw their money from the banks at the same time, causing bank runs. These problems would not have happened to Maker because of its emergency shutdown mechanism, which would ensure that all users can withdraw their assets. Maker has no power to loan or take away a user's assets.

The Federal Reserve has served its purpose in America for a long time, but it is an old centralized finance system, while MakerDAO is a new decentralized finance system. As MakerDAO continues to constantly change and improve, it can act as a stable, comprehensive medium of exchange.

[Insert Figure 7 here.]

7. **CONCLUSION**

Decentralized finance constitutes a burgeoning field at the intersection of finance and technology. Due to its nascency, progress occurs at a rapid pace rendering it difficult to reliably forecast its future state even in the near to medium term. While the verdict on blockchain technologies "revolutionizing" finance is yet out, its implications for practice and regulation are at the forefront of current issues. Across the board, whether for payment systems, securities settlement, or payment flows in the interbank market, blockchain technologies are considered as replacement for legacy systems. In many cases, private proposals, such as Maker, compete with those of the public sector. Our understanding of those alternative solutions will shape the direction in which decentralized finance evolves.

**FIGURES**

Figure 1: The Maker Logo. Source: https://globalcrypto.tv/maker-dao-and-collateralised-loans/maker-dao-logo/

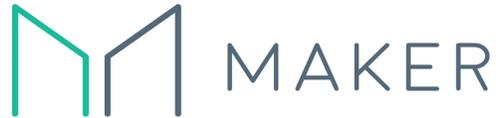

Figure 2: The Dai Logo. Source: https://www.pngitem.com/middle/hhxhTim_dai-logo-png-transparent-png/

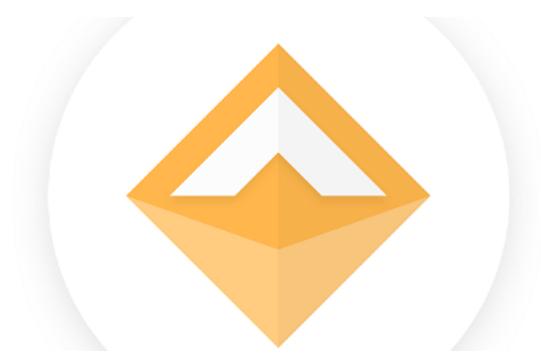

Figure 3: MakerDAO Total Value Locked (TVL). The graph shows a low level of TVL prior to year 2020, a rapid rise starting near the beginning of 2021, and a crash-like decline in 2022. Source: https://defillama.com/protocol/makerdao

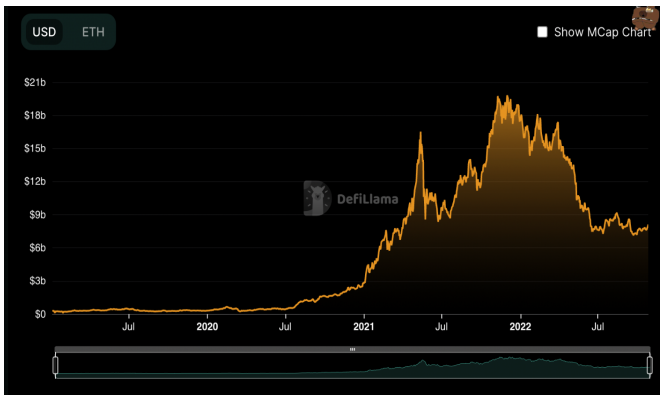

Figure 4: Expanded Graph of MakerDAO's Total Value Locked (TVL) in 2022. Maker TVL declined from $17.5 billion in January to less than $9 billion in October. Source: https://defillama.com/protocol/makerdao

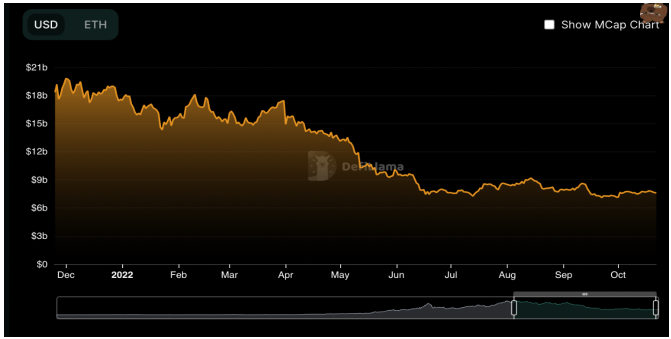

Figure 5: Total Value Locked (TVL) of Ethereum. Source: https://defillama.com/protocol/makerdao

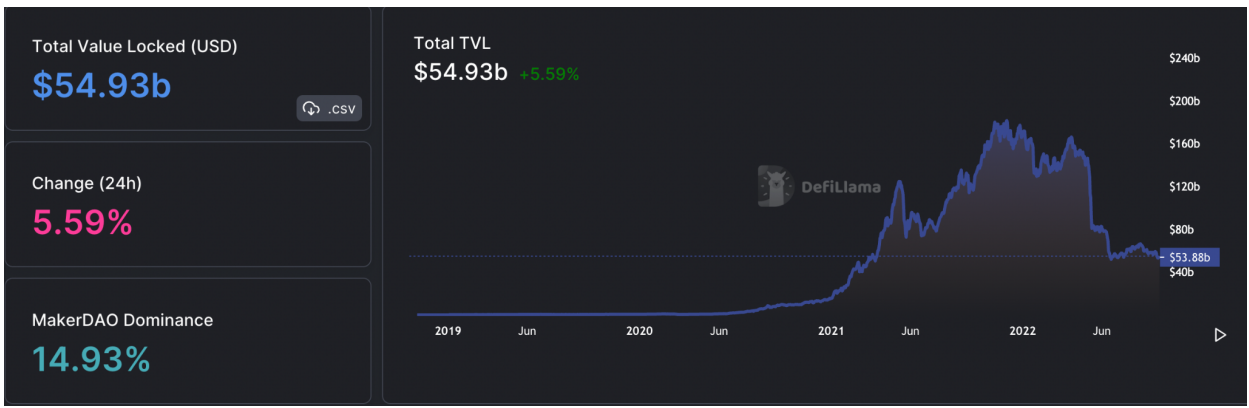

Figure 6: Comparisons between the Traditional, Centralized Financial System and the Decentralized Financial System. Source: https://medium.com/stably-blog/decentralized-finance-vs-traditional-finance-what-you-need-to-know-3b57aed7a0c2

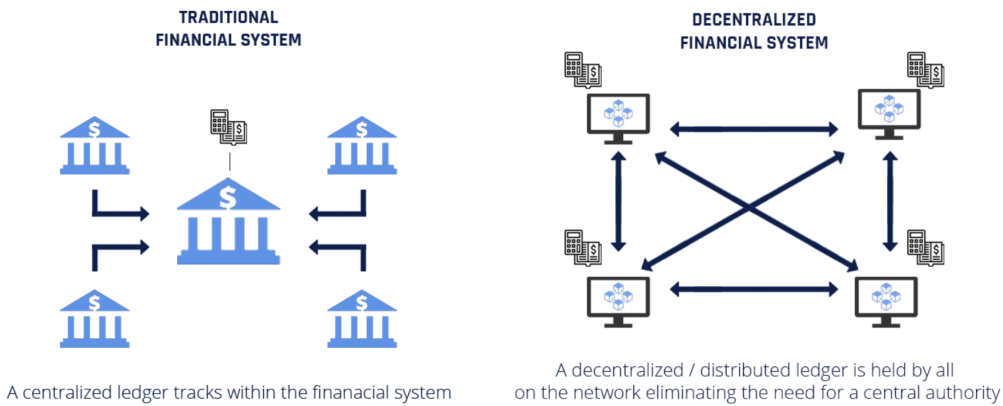

Figure 7: From Sender to Receiver: Comparisons between a Traditional Financial System and the Decentralized Financial System. Source: https://medium.com/tchyon/defi-a-new-chapter-in-indias-finance-book-e7b653683bbf.

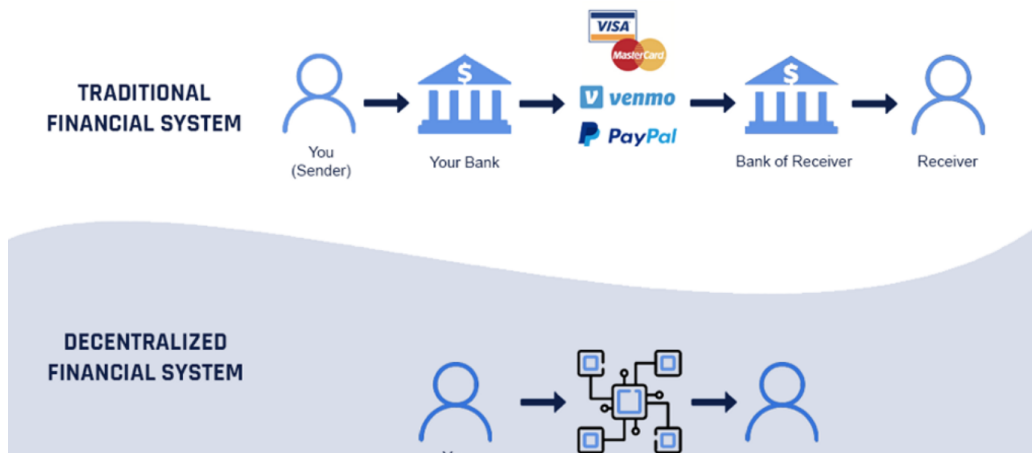